# The superposition invariance of unitary operators and maximally entangled state

Xin-Wei Zha, Yun-Guang Zhang and Jian-Xia Qi

School of Science, Xi'an University of Posts and Telecommunications, Xi'an, 710121, P R China

**Abstract**

In this paper, we study the superposition invariance of unitary operators and maximally entangled state respectively. Furthermore, we discuss the set of orthogonal maximally entangled states. We find that orthogonal basis of maximally entangled states can be divided into k subspaces. It is shown that some entanglement properties of superposed state in every subspace are invariant.

## I. Introduction

Entanglement is considered to be the most important resource for quantum information and computation. It is, therefore, essential to exploit entangled states and reveal their entanglement properties with respect to the usefulness for given quantum information tasks [1–3]. Non-locality is, on the other hand, one of the astonishing phenomena in quantum mechanics[4–5].In recent years，entanglement has a great many perspective applications in quantum computing and quantum information. Recently a new question has been raised concerning of the entanglement of the superposed states [6–13]. It is known that the entanglement property may change drastically under superposition. For example in Ref [13], superposition of two product states may yield a GHZ state and, on the contrary, superposition of two GHZ states may lead to a product state. Preeti Parashar and Swapan Ranat[14]calculated the analytic expression for geometric measure of entanglement for arbitrary superposition of two N-qubit canonical orthogonal Greenberger-Horne-Zeilinger(GHZ) states and the same for two W states. Seyed Javad Akhtarshenas [15] investigated lower and upper bounds on the concurrence of the superposition of two states.

For a state constituted by several superposed states, what are its quantum entanglement properties? How is it related to the several constituents and to the coefficients? Whether can we find that entanglement properties are invariant for a state constituted by several superposed states? In this paper, we will explicitly calculate the the superposition of unitary operators. Furthermore, we analyze the entanglement properties of the superposed state of orthogonal maximally entangled states.

## 2. The superposition invariance of unitary operators

Let us consider operators

$$X=\begin{pmatrix} 0 & 1 \\ 1 & 0 \end{pmatrix}, Y=\begin{pmatrix} 0 & 1 \\ -1 & 0 \end{pmatrix}, Z=\begin{pmatrix} 1 & 0 \\ 0 & -1 \end{pmatrix}, I=\begin{pmatrix} 1 & 0 \\ 0 & 1 \end{pmatrix}. \quad (1)$$

Is is easy to show those operators $X$, $Y$, $Z$, $I$ are unitary operators.

Now let us look at $A=c_1X+c_2Z = \begin{pmatrix} c_2 & c_1 \\ c_1 & -c_2 \end{pmatrix}$, (2)

If superposition coefficients $c_1$, $c_2$ are real numbers and $c_1^2 + c_2^2 = 1$.

It is easy to show that operator $A$ is also a unitary operator.

$$\text{Similarly, if } B=c_1Y+c_2I = \begin{pmatrix} c_2 & c_1 \\ -c_1 & c_2 \end{pmatrix}, \quad (3)$$

We can show that operator $B$ is also a unitary operator.

Similarly, we know that

$$X_1=\begin{pmatrix} 0 & 1 \\ 1 & 0 \end{pmatrix}, Y_1=\begin{pmatrix} 0 & 1 \\ -1 & 0 \end{pmatrix}, Z_1=\begin{pmatrix} 1 & 0 \\ 0 & -1 \end{pmatrix}, I_1=\begin{pmatrix} 1 & 0 \\ 0 & 1 \end{pmatrix}.$$

$$X_2=\begin{pmatrix} 0 & 1 \\ 1 & 0 \end{pmatrix}, Y_2=\begin{pmatrix} 0 & 1 \\ -1 & 0 \end{pmatrix}, Z_2=\begin{pmatrix} 1 & 0 \\ 0 & -1 \end{pmatrix}, I_2=\begin{pmatrix} 1 & 0 \\ 0 & 1 \end{pmatrix} \quad (4)$$

Then we obtain

$$I_1 \otimes I_2 = \begin{pmatrix} 1 & 0 & 0 & 0 \\ 0 & 1 & 0 & 0 \\ 0 & 0 & 1 & 0 \\ 0 & 0 & 0 & 1 \end{pmatrix}, \quad I_1 \otimes X_2 = \begin{pmatrix} 0 & 1 & 0 & 0 \\ 1 & 0 & 0 & 0 \\ 0 & 0 & 0 & 1 \\ 0 & 0 & 1 & 0 \end{pmatrix}$$

$$I_1 \otimes Y_2 = \begin{pmatrix} 0 & 1 & 0 & 0 \\ -1 & 0 & 0 & 0 \\ 0 & 0 & 0 & 1 \\ 0 & 0 & -1 & 0 \end{pmatrix}, \quad I_1 \otimes Z_2 = \begin{pmatrix} 1 & 0 & 0 & 0 \\ 0 & -1 & 0 & 0 \\ 0 & 0 & 1 & 0 \\ 0 & 0 & 0 & -1 \end{pmatrix}$$

$$X_1 \otimes I_2 = \begin{pmatrix} 0 & 0 & 1 & 0 \\ 0 & 0 & 0 & 1 \\ 1 & 0 & 0 & 0 \\ 0 & 1 & 0 & 0 \end{pmatrix}, \quad X_1 \otimes X_2 = \begin{pmatrix} 0 & 0 & 0 & 1 \\ 0 & 0 & 1 & 0 \\ 0 & 1 & 0 & 0 \\ 1 & 0 & 0 & 0 \end{pmatrix}$$

$$X_1 \otimes Y_2 = \begin{pmatrix} 0 & 0 & 0 & 1 \\ 0 & 0 & -1 & 0 \\ 0 & 1 & 0 & 0 \\ -1 & 0 & 0 & 0 \end{pmatrix}, \quad X_1 \otimes Z_2 = \begin{pmatrix} 0 & 0 & 1 & 0 \\ 0 & 0 & 0 & -1 \\ 1 & 0 & 0 & 0 \\ 0 & -1 & 0 & 0 \end{pmatrix}$$

$$Y_1 \otimes I_2 = \begin{pmatrix} 0 & 0 & 1 & 0 \\ 0 & 0 & 0 & 1 \\ -1 & 0 & 0 & 0 \\ 0 & -1 & 0 & 0 \end{pmatrix}, \quad Y_1 \otimes X_2 = \begin{pmatrix} 0 & 0 & 0 & 1 \\ 0 & 0 & 1 & 0 \\ 0 & -1 & 0 & 0 \\ -1 & 0 & 0 & 0 \end{pmatrix}$$

$$Y_1 \otimes Y_2 = \begin{pmatrix} 0 & 0 & 0 & 1 \\ 0 & 0 & -1 & 0 \\ 0 & -1 & 0 & 0 \\ 1 & 0 & 0 & 0 \end{pmatrix}, \quad Y_1 \otimes Z_2 = \begin{pmatrix} 0 & 0 & 1 & 0 \\ 0 & 0 & 0 & -1 \\ -1 & 0 & 0 & 0 \\ 0 & 1 & 0 & 0 \end{pmatrix}$$

$$Z_1 \otimes I_2 = \begin{pmatrix} 1 & 0 & 0 & 0 \\ 0 & 1 & 0 & 0 \\ 0 & 0 & -1 & 0 \\ 0 & 0 & 0 & -1 \end{pmatrix}, \quad Z_1 \otimes X_2 = \begin{pmatrix} 0 & 1 & 0 & 0 \\ 1 & 0 & 0 & 0 \\ 0 & 0 & 0 & -1 \\ 0 & 0 & -1 & 0 \end{pmatrix}$$

$$Z_1 \otimes Y_2 = \begin{pmatrix} 0 & 1 & 0 & 0 \\ -1 & 0 & 0 & 0 \\ 0 & 0 & 0 & -1 \\ 0 & 0 & 1 & 0 \end{pmatrix}, \quad Z_1 \otimes Z_2 = \begin{pmatrix} 1 & 0 & 0 & 0 \\ 0 & -1 & 0 & 0 \\ 0 & 0 & -1 & 0 \\ 0 & 0 & 0 & 1 \end{pmatrix} \quad (5)$$

We know that operators $I_1 \otimes I_2$, $I_1 \otimes X_2$, $I_1 \otimes Y_2$, $I_1 \otimes Z_2$, $X_1 \otimes I_2$, $X_1 \otimes X_2$, $X_1 \otimes Y_2$, $X_1 \otimes Z_2$, $Y_1 \otimes I_2$, $Y_1 \otimes X_2$, $Y_1 \otimes Y_2$, $Y_1 \otimes Z_2$, $Z_1 \otimes I_2$, $Z_1 \otimes X_2$, $Z_1 \otimes Y_2$, $Z_1 \otimes Z_2$ are unitary operators.

Now let us look at

$$U_{12}^1 = c_1 I_1 \otimes I_2 + c_2 I_1 \otimes Y_2 + c_3 Y_1 \otimes X_2 + c_4 Y_1 \otimes Z_2$$

$$= \begin{pmatrix} c_1 & c_2 & c_4 & c_3 \\ -c_2 & c_1 & c_3 & -c_4 \\ -c_4 & -c_3 & c_1 & c_2 \\ -c_3 & c_4 & -c_2 & c_1 \end{pmatrix}, \quad (6)$$

Assume superposition coefficients $c_1$, $c_2, c_3$, $c_4$ are real numbers and $c_1^2 + c_2^2 + c_3^2 + c_4^2 = 1$.

It is easy to show that operator $U_{12}^1$ is a unitary operator.

Similarly, if

$$U_{12}^2 = c_1 I_1 \otimes Z_2 + c_2 I_1 \otimes X_2 + c_3 Y_1 \otimes Y_2 + c_4 Y_1 \otimes I_2 = \begin{pmatrix} c_1 & c_2 & c_4 & c_3 \\ c_2 & -c_1 & -c_3 & c_4 \\ -c_4 & -c_3 & c_1 & c_2 \\ c_3 & -c_4 & c_2 & -c_1 \end{pmatrix} \quad (7)$$

$$U_{12}^3 = c_1 Z_1 \otimes Z_2 + c_2 Z_1 \otimes X_2 + c_3 X_1 \otimes Y_2 + c_4 X_1 \otimes I_2 = \begin{pmatrix} c_1 & c_2 & c_4 & c_3 \\ c_2 & -c_1 & -c_3 & c_4 \\ c_4 & c_3 & -c_1 & -c_2 \\ -c_3 & c_4 & -c_2 & c_1 \end{pmatrix} \quad (8)$$

$$U_{12}^4 = c_1 Z_1 \otimes I_2 + c_2 Z_1 \otimes Y_2 + c_3 X_1 \otimes X_2 + c_4 X_1 \otimes Z_2 = \begin{pmatrix} c_1 & c_2 & c_4 & c_3 \\ -c_2 & c_1 & c_3 & -c_4 \\ c_4 & c_3 & -c_1 & -c_2 \\ c_3 & -c_4 & c_2 & -c_1 \end{pmatrix} \quad (9)$$

It is easy to show that operators $U_{12}^2, U_{12}^3, U_{12}^4$ are also unitary operators.

It can be show that $U_{12}^i \neq U_1 \otimes U_2, i = 1, 2, 3, 4$, i.e. those operators cannot transfer into two unitary operator products.

But for some special cases, for example, if $c_3 = c_4 = 0$, we have

$$U_{12}^1 = c_1 I_1 \otimes I_2 + c_2 I_1 \otimes Y_2 = I_1 \otimes (c_1 I_2 + c_2 Y_2) = I_1 \otimes B_2$$
$$U_{12}^2 = c_1 I_1 \otimes Z_2 + c_2 I_1 \otimes X_2 = I_1 \otimes (c_1 Z_2 + c_2 X_2) = I_1 \otimes A_2$$
$$U_{12}^3 = c_1 Z_1 \otimes Z_2 + c_2 Z_1 \otimes X_2 = Z_1 \otimes (c_1 Z_2 + c_2 X_2) = Z_1 \otimes A_2$$
$$U_{12}^4 = c_1 Z_1 \otimes I_2 + c_2 Z_1 \otimes Y_2 = Z_1 \otimes (c_1 I_2 + c_2 Y_2) = Z_1 \otimes B_2 \quad (10)$$

if $c_1 = c_2 = 0$, we have

$$U_{12}^1 = c_3 Y_1 \otimes X_2 + c_4 Y_1 \otimes Z_2 = Y_1 \otimes A_2$$

$$U_{12}^2 = c_3 Y_1 \otimes Y_2 + c_4 Y_1 \otimes I_2 = Y_1 \otimes B_2,$$

$$U_{12}^3 = c_3 X_1 \otimes Y_2 + c_4 X_1 \otimes I_2 = X_1 \otimes B_2$$

$$U_{12}^4 = c_3 X_1 \otimes X_2 + c_4 X_1 \otimes Z_2 = X_1 \otimes A_2 \quad (11)$$

In this case, superposition operators are not only unitary operator, but also two unitary operator products.

## 3 The superposition invariance of maximally entangled basis

Let us assume that $\{|\varphi^1\rangle, |\varphi^2\rangle \cdots, |\varphi^n\rangle\}$, are a set of complete orthogonal maximally entangled states that constitute the orthogonal basis. These states can be divided into k sets $\{|\varphi_k^1\rangle, |\varphi_k^2\rangle \cdots, |\varphi_k^m\rangle\}, k = 1, 2 \cdots l$, certainly, $ml = n$,

Where $\langle \varphi_k^i | \varphi_l^j \rangle = \delta_{kl}\delta_{ij}$, and

$$|\varphi_k^i\rangle = U_1 \otimes U_2 \otimes \cdots \otimes U_n |\varphi_1^i\rangle, i = 1, 2, \cdots m \quad (12)$$

And $U_j$ is a unitary operation acting on the Hilbert space of the jth party.

In subspace, we find that some entanglement properties of superposed state

$$|\psi\rangle = c_1|\varphi_k^1\rangle + c_2|\varphi_k^2\rangle + \cdots + c_m|\varphi_k^m\rangle \quad (13)$$

is invariant.

3.1 For two-qubit system, it is well known that there are four orthogonal maximally entangled states, i.e., four Bell states,

$$|\psi\rangle^+ = \frac{1}{\sqrt{2}}(|01\rangle + |10\rangle),$$

$$|\psi\rangle^- = \frac{1}{\sqrt{2}}(|01\rangle - |10\rangle),$$

$$|\varphi\rangle^+ = \frac{1}{\sqrt{2}}(|00\rangle + |11\rangle), \quad (14)$$

$$|\varphi\rangle^- = \frac{1}{\sqrt{2}}(|00\rangle - |11\rangle)$$

On the other hand, we know that

$$|\varphi\rangle^+ = I_1|\varphi\rangle^+, |\varphi\rangle^- = Z_1|\varphi\rangle^+, |\psi\rangle^+ = X_1|\varphi\rangle^+, |\psi\rangle^- = Y_1|\varphi\rangle^+;$$

Let us express

$$\{|\varphi_1^1\rangle, |\varphi_1^2\rangle\} = \{I_1|\varphi\rangle^+, Y_1|\varphi\rangle^+\} = \{|\varphi\rangle^+, |\psi\rangle^-\},$$

$$\{|\varphi_2^1\rangle, |\varphi_2^2\rangle\} = \{Z_1|\varphi\rangle^+, X_1|\varphi\rangle^+\} = \{|\varphi\rangle^-, |\psi\rangle^+\}$$

The superposition of two states in the same sets is

$$|\psi\rangle_s = c_1|\varphi_k^1\rangle + c_2|\varphi_k^2\rangle, k = 1, 2 \quad (15)$$

For $k = 1$, the superposition state

$$|\psi\rangle_s = c_1|\varphi\rangle^+ + c_2|\psi\rangle^- \quad (16)$$

We have $|\psi\rangle_s = c_1I_1 + c_2Y_1|\varphi\rangle^+$, By using Eq (3), it is not difficult to find that the superposition states is also a maximally entangled state.

Similarly, the superposition state $|\psi\rangle_s = c_1|\varphi\rangle^- + c_2|\psi\rangle^+$ is also a maximally entangled state.

3.2  For three-qubit system, there are eight orthogonal maximally entangled states, i.e., eight GHZ states,

$$|\psi_1\rangle_{123} = \frac{1}{\sqrt{2}}(|000\rangle + |111\rangle)_{123}$$

$$|\psi_2\rangle_{123} = \frac{1}{\sqrt{2}}(|000\rangle - |111\rangle)_{123}$$

$$|\psi_3\rangle_{123} = \frac{1}{\sqrt{2}}(|001\rangle + |110\rangle)_{123}$$

$$|\psi_4\rangle_{123} = \frac{1}{\sqrt{2}}(|001\rangle - |110\rangle)_{123}$$

$$|\psi_5\rangle_{123} = \frac{1}{\sqrt{2}}(|010\rangle + |101\rangle)_{123}$$

$$|\psi_6\rangle_{123} = \frac{1}{\sqrt{2}}(|010\rangle - |101\rangle)_{123}$$

$$|\psi_7\rangle_{123} = \frac{1}{\sqrt{2}}(|100\rangle + |011\rangle)_{123}$$

$$|\psi_8\rangle_{123} = \frac{1}{\sqrt{2}}(|100\rangle - |011\rangle)_{123} \qquad (17)$$

On the other hand, we know that $|\psi_1\rangle_{123} = I_1|\psi_1\rangle_{123}$, $|\psi_2\rangle_{123} = Z_1|\psi_1\rangle_{123}$, $|\psi_3\rangle_{123} = X_3|\psi_1\rangle_{123}$, $|\psi_4\rangle_{123} = -Y_3|\psi_1\rangle_{123}$ ; $|\psi_5\rangle_{123} = X_2|\psi_1\rangle_{123}$, $|\psi_6\rangle_{123} = -Y_2|\psi_1\rangle_{123}$, $|\psi_7\rangle_{123} = X_1|\psi_1\rangle_{123}$, $|\psi_8\rangle_{123} = -Y_1|\psi_1\rangle_{123}$.

Let us express

$\{|\varphi_1^1\rangle, |\varphi_1^2\rangle\} = \{|\psi_1\rangle_{123}, |\psi_4\rangle_{123}\}$,

$\{|\varphi_2^1\rangle, |\varphi_2^2\rangle\} = \{|\psi_2\rangle_{123}, |\psi_3\rangle_{123}\}$

$\{|\varphi_3^1\rangle, |\varphi_3^2\rangle\} = \{|\psi_5\rangle_{123}, |\psi_8\rangle_{123}\}$

$\{|\varphi_4^1\rangle, |\varphi_4^2\rangle\} = \{|\psi_6\rangle_{123}, |\psi_7\rangle_{123}\}$,

We have $|\varphi_2^i\rangle = \sigma_{1z}|\varphi_1^i\rangle$, $|\varphi_3^i\rangle = \sigma_{2x}|\varphi_1^i\rangle$, $|\varphi_4^i\rangle = \sigma_{1z}\sigma_{2x}|\varphi_1^i\rangle$.

Obviously, the superposition of two states in the same sets is

$$|\psi\rangle_s = c_1|\varphi_k^1\rangle + c_2|\varphi_k^2\rangle, k = 1, 2, 3, 4. \qquad (18)$$

And which is also a maximally entangled state.

3.3 For four-qubit system, there are sixteen orthogonal maximally entangled states, i.e., sixteen cluster states [16]. These states can be divided into 4 sets $\{|\varphi_k^1\rangle, |\varphi_k^2\rangle, |\varphi_k^3\rangle, |\varphi_k^4\rangle\}, k=1,2,3,4$, where $\langle \varphi_k^i | \varphi_l^j \rangle = \delta_{kl}\delta_{ij}$.

We have

$$\begin{cases} |\varphi_1^1\rangle = \frac{1}{2}(|0000\rangle + |0101\rangle + |1010\rangle - |1111\rangle)_{1234} \\ |\varphi_1^2\rangle = \frac{1}{2}(-|0001\rangle + |0100\rangle + |1011\rangle + |1110\rangle)_{1234} \\ |\varphi_1^3\rangle = \frac{1}{2}(|0010\rangle + |0111\rangle - |1000\rangle + |1101\rangle)_{1234} \\ |\varphi_1^4\rangle = \frac{1}{2}(|0011\rangle - |0110\rangle + |1001\rangle + |1100\rangle)_{1234} \end{cases} \quad (19)$$

And

$$|\varphi_2^i\rangle = \sigma_{1z}|\varphi_1^i\rangle, \ |\varphi_3^i\rangle = \sigma_{1x}|\varphi_1^i\rangle, \ |\varphi_4^i\rangle = \sigma_{2x}|\varphi_1^i\rangle \quad (20)$$

It is easy to show that $\pi_{12} = Tr_{12}\rho_{12}^2 = \frac{1}{4}$ for sixteen cluster states.

The superposition states in the same sets is

$$|\psi\rangle_s = c_1|\varphi_1^1\rangle + c_2|\varphi_1^2\rangle + c_3|\varphi_1^3\rangle + c_4|\varphi_1^4\rangle. \quad (21)$$

Using Esq.(6-8), one can check that

$$\pi_{12} = Tr_{12}\rho_{12}^2 = \frac{1}{4}. \quad (22)$$

Therefore, the entanglement property of superposed state $\pi_{12}$ is invariant.

3.4 For a maximally entangled five-qubit state, Brown *et al.* [17] have achieved it. And it can be expressed as

$$|\psi\rangle_{12345} = \frac{1}{2}(|001\rangle|\phi_-\rangle + |010\rangle|\psi_-\rangle + |100\rangle|\phi_+\rangle + |111\rangle|\psi_+\rangle)_{12345}, \quad (23a)$$

Where

$|\psi_\pm\rangle = \frac{1}{\sqrt{2}}(|00\rangle \pm |11\rangle)$, and $|\phi_\pm\rangle = \frac{1}{\sqrt{2}}(|01\rangle \pm |10\rangle)$ are Bell states.

Eq. (23a) can be rewritten as

$$|\psi\rangle_{12345} = \frac{1}{2\sqrt{2}}(|00101\rangle - |00110\rangle + |01000\rangle - |01011\rangle$$
$$+ |10001\rangle + |10010\rangle + |11100\rangle + |11111\rangle)_{12345} \quad (23b)$$

For five-qubit, there are 32 orthogonal maximally entangled states,

$$\begin{cases} |\varphi_1^1\rangle = \frac{1}{2\sqrt{2}}(|00101\rangle - |00110\rangle + |01000\rangle - |01011\rangle \\ \qquad\qquad + |10001\rangle + |10010\rangle + |11100\rangle + |11111\rangle)_{12345} \\ \\ |\varphi_1^2\rangle = \frac{1}{2\sqrt{2}}(|00001\rangle + |00010\rangle - |01100\rangle - |01111\rangle \\ \qquad\qquad - |10101\rangle + |10110\rangle + |11000\rangle - |11011\rangle)_{12345} \\ \\ |\varphi_1^3\rangle = \frac{1}{2\sqrt{2}}(|00000\rangle - |00011\rangle - |01101\rangle + |01110\rangle \\ \qquad\qquad + |10100\rangle + |10111\rangle - |11001\rangle - |11010\rangle)_{12345} \\ \\ |\varphi_1^4\rangle = \frac{1}{2\sqrt{2}}(|00100\rangle + |00111\rangle + |01001\rangle + |01010\rangle \\ \qquad\qquad - |10000\rangle + |10011\rangle - |11101\rangle + |11110\rangle)_{12345} \end{cases} \quad (24)$$

And

$$\{|\varphi_2^1\rangle, |\varphi_2^2\rangle, |\varphi_2^3\rangle, |\varphi_2^4\rangle\} = \hat{\sigma}_{1z}\{|\varphi_1^1\rangle, |\varphi_1^2\rangle, |\varphi_1^3\rangle, |\varphi_1^4\rangle\},$$

$$\{|\varphi_3^1\rangle, |\varphi_3^2\rangle, |\varphi_3^3\rangle, |\varphi_3^4\rangle\} = \hat{\sigma}_{2z}\{|\varphi_1^1\rangle, |\varphi_1^2\rangle, |\varphi_1^3\rangle, |\varphi_1^4\rangle\},$$

$$\{|\varphi_4^1\rangle, |\varphi_4^2\rangle, |\varphi_4^3\rangle, |\varphi_4^4\rangle\} = \hat{\sigma}_{1z}\hat{\sigma}_{2z}\{|\varphi_1^1\rangle, |\varphi_1^2\rangle, |\varphi_1^3\rangle, |\varphi_1^4\rangle\},$$

$$\{|\varphi_5^1\rangle, |\varphi_5^2\rangle, |\varphi_5^3\rangle, |\varphi_5^4\rangle\} = \hat{\sigma}_{4z}\{|\varphi_1^1\rangle, |\varphi_1^2\rangle, |\varphi_1^3\rangle, |\varphi_1^4\rangle\},$$

$$\{|\varphi_6^1\rangle, |\varphi_6^2\rangle, |\varphi_6^3\rangle, |\varphi_6^4\rangle\} = \hat{\sigma}_{5z}\{|\varphi_1^1\rangle, |\varphi_1^2\rangle, |\varphi_1^3\rangle, |\varphi_1^4\rangle\},$$

$$\{|\varphi_7^1\rangle, |\varphi_7^2\rangle, |\varphi_7^3\rangle, |\varphi_7^4\rangle\} = \hat{\sigma}_{1z}\hat{\sigma}_{4z}\{|\varphi_1^1\rangle, |\varphi_1^2\rangle, |\varphi_1^3\rangle, |\varphi_1^4\rangle\},$$

$$\{|\varphi_8^1\rangle, |\varphi_8^2\rangle, |\varphi_8^3\rangle, |\varphi_8^4\rangle\} = \hat{\sigma}_{1z}\hat{\sigma}_{5z}\{|\varphi_1^1\rangle, |\varphi_1^2\rangle, |\varphi_1^3\rangle, |\varphi_1^4\rangle\}. \quad (25)$$

It is easy to show that $\pi_{12} = Tr_{12}\rho_{12}^2 = \frac{1}{4}$ for 32 orthogonal maximally entangled states.

Assuming that the superposition states in the same sets is

$$|\psi\rangle_s = c_1|\varphi_1^1\rangle + c_2|\varphi_1^2\rangle + c_3|\varphi_1^3\rangle + c_4|\varphi_1^4\rangle . \quad (26)$$

By simple algebra we get

$$\pi_{12} = Tr_{12}\rho_{12}^2 = \frac{1}{4} \quad (27)$$

Therefore, the entanglement property of superposed state $\pi_{12}$ is also invariant.

## 4. Discussions and conclusion

In summary, we introduce the invariant of superposition of unitary operators. For 2, 3, 4, 5-qubit, we introduce the complete set of orthogonal maximally entangled states. We find that orthogonal basis with the maximally entangled states can be divided into k subspace. We have discussed the entanglement properties of the superposed state in every subspace. It is shown that some entanglement properties of superposed state in every subspace are invariant. As is shown in reference [18], we know that if $\pi_{ij} = Tr_{ij}\rho_{ij}^2 = \frac{1}{4}$, using this state as quantum channel in teleportation, the unknown two particle state can be teleported perfectly. We hope that maximally entangled basis to be proved as useful to quantum information theory.

Acknowledgements

This work is supported by Shaanxi Natural Science Foundation under Contract (No. 2013JM1009 and 2015JM6263 ).

Reference

[1] Horodecki R, Horodecki P, Horodecki M, and Horodecki K 2009 *Rev. Mod. Phys.* 81 865 .

[2] Lo Franco R, D'Arrigo A, Falci G, Compagno G, and Paladino E 2012 *Phys. Scr.* T147 014019.

[3] Bellomo B, Lo Franco R, Maniscalco S, and Compagno G 2010 *Phys. Scr.* T140 014014.

[4] Lo Franco R, Compagno G, Messina A, and Napoli A 2005 *Phys. Rev. A* 72 053806.
[5] Acin A, Gisin N, and Masanes L 2006 *Phys. Rev. Lett.* 97 120405.
[6] Linden N, Popescu S and Smolin J A 2006 *Phys. Rev. Lett.* 97, 100502.
[7] Yu C S, Yi X X and Song H S 2007 *Phys. Rev. A* 75, 022332.

[8] Ou Y C and Fan H 2007 *Phys. Rev. A* 76, 022320.

[9] Cavalvanti D, Terra Cunha M O andAcin A 2007 *Phys. Rev. A* 76, 042329.

[10] Gour G 2007 *Phys.Rev.A* 76 052320.

[11] Song W, Liu N L and Chen Z B 2007 *Phys. Rev. A* 76, 054303.

[12] Zhang S, Zhou Z W and Guo G G 2009 *Chin. Phys. Lett.* 26 020304 .

[13] Zhang D H, Zhou D L and FAN H 2010 *Chin. Phys. Lett.* 27 090306 .

[14] Parashar P and Rana S 2011 *Phys. Rev. A* 83  032301.

[15] Akhtarshenas S J 2011 *Phys. Rev. A.* 83, 042306.

[16] Zha X W, Song H Y, and Feng F 2011 Commun. Theor. Phys. 56 827

[17] Brown I D K, Stepney S, Sudbery A and Braunstein S L 2005 J. Phys. A: Math. Gen. 38, 1119 .

[18] Zha X W, Song H Yand Ren K F 2010 IJQI 8(8) 1251 .